\documentclass[prb,aps,superscriptaddress,twocolumn]{revtex4}

\usepackage{graphicx}
\usepackage{amssymb}
\usepackage{times}
\usepackage{amsmath}
\usepackage{dcolumn}
\usepackage{bm}
\usepackage{colordvi}
\usepackage{color}
\usepackage{epsfig}

\newcommand{\Av}{{\textrm{V,A}}}

\newcommand{\Bv}{{\textrm{V,B}}}

\newcommand{\spn}{{\sigma_\mathrm{pn}}}

\begin{document}

\title{Preparation of distilled and purified continuous variable entangled states}

\author{Boris Hage}
\affiliation{Albert-Einstein-Institut, Max-Planck-Institut f\"ur Gravitationsphysik and Leibniz Universit\"at Hannover, Callinstr. 38, 30167 Hannover, Germany}

\author{Aiko Samblowski}
\affiliation{Albert-Einstein-Institut, Max-Planck-Institut f\"ur Gravitationsphysik and Leibniz Universit\"at Hannover, Callinstr. 38, 30167 Hannover, Germany}

\author{James DiGuglielmo}
\affiliation{Albert-Einstein-Institut, Max-Planck-Institut f\"ur Gravitationsphysik and Leibniz Universit\"at Hannover, Callinstr. 38, 30167 Hannover, Germany}

\author{Alexander Franzen}
\affiliation{Albert-Einstein-Institut, Max-Planck-Institut f\"ur Gravitationsphysik and Leibniz Universit\"at Hannover, Callinstr. 38, 30167 Hannover, Germany}

\author{Jarom\'{i}r Fiur\'{a}\v{s}ek}
\affiliation{Department of Optics, Palack\'y University, 17. listopadu 50, 77200 Olomouc,
Czech Republic}

\author{Roman Schnabel*}
\affiliation{Albert-Einstein-Institut, Max-Planck-Institut f\"ur Gravitationsphysik and Leibniz Universit\"at Hannover, Callinstr. 38, 30167 Hannover, Germany}

\date{\today}
\maketitle

\textbf{The distribution of entangled quantum states of light over
long distances is a major challenge in the field of quantum
information. Optical losses, phase diffusion and mixing with thermal
states lead to decoherence and destroy the nonclassical states after
some finite transmission line length. Quantum repeater
protocols\cite{Briegel98,Duan01} combining  quantum
memory\cite{Julsgaard04}, entanglement
distillation\cite{Bennett96,Deutsch96} and entanglement
swapping\cite{Zukowski93} were proposed to overcome this problem.
Here  we report on the first experimental demonstration of
entanglement distillation in the continuous variable
regime\cite{Browne03,Eisert04,Fiurasek07}.
 Entangled squeezed states were
first disturbed by random phase fluctuations and then distilled and
purified using interference on beam splitters and homodyne
detection. Measurements of covariance matrices clearly indicated a
regained strength of entanglement and purity of the distilled state.
Contrasting previous
demonstrations in the complementary discrete variable
regime\cite{Pan03,Zhao03}, our scheme achieved the actual
preparation of the distilled states, which might therefore be used
to improve the quality of down-stream applications
such as quantum teleportation\cite{Furusawa98}.}


Quantum information makes use of the special properties of quantum
states in order to improve the quality of  communication and
information processing tasks. Generally, a quantized field can be
described by the number operator or alternatively by two non-commuting position and
momentum-like operators. The corresponding measurement results have
either discrete or continuous spectra and form the basis of discrete
variable (DV) or continuous variable (CV) quantum information,
respectively.
In the regime of continuous variables, entangled states of light can
be deterministically generated in optical parametric amplifiers
(OPAs), precisely manipulated with linear optics, and measured with
very high efficiency in balanced homodyne detectors. These entangled
two-mode squeezed states show Gaussian probability distributions and
were utilized for quantum teleportation\cite{Furusawa98} and
entanglement swapping\cite{TYAF05,Jia04}. Entangled states of the
collective spins of two atomic ensembles analogous to two-mode
squeezed states have been generated\cite{Julsgaard01}, storage of
quantum states of light in an atomic memory has been
demonstrated\cite{Julsgaard04} and teleportation from light onto an
atomic ensemble has been reported\cite{Sherson06}. High-speed
quantum cryptography with coherent light beams and homodyne
detection has been demonstrated\cite{Grosshans03}. All these
spectacular achievements reveal the great potential of this approach
to quantum information processing.

\begin{figure}[!t!]
\includegraphics[width=0.85\linewidth]{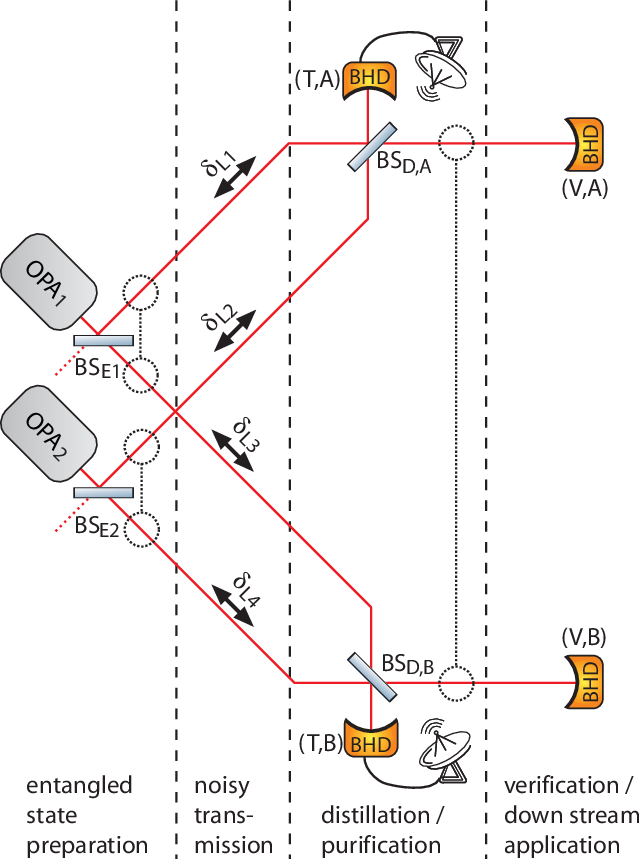}
\caption{\textbf{Experimental setup.} OPA: optical parametric amplifier, BS$_{\mathrm{D}}$:
distillation beam splitter, BS$_{\mathrm{E}}$: entangling beam splitter, BHD:
balanced homodyne detector. $\delta_{\mathrm{Li}}$ indicate
independently fluctuating optical path lengths. The combination of both BHD$_{\mathrm{T,A/B}}$ provided the trigger signal for successful distillation, whereas BHD$_{\mathrm{V,A/B}}$ were used for its independent verification. The latter were not necessary for the distillation protocol to work.}
\label{setup}
\end{figure}

A missing piece in this toolbox has been a feasible protocol for
entanglement distillation and purification.
Entanglement distillation\cite{Bennett96,Deutsch96} extracts from
several shared copies of weakly entangled mixed states a single copy
of a highly entangled state using only local quantum operations and
classical communication between the two parties sharing the states.
This turned out to be a very challenging task for CV states since it
was proved that it is impossible to distill Gaussian entangled states by means of the
experimentally accessible Gaussian operations\cite{Eisert02,Giedke02}.
However, a whole class of important decoherence processes give rise to non-Gaussian noise and
therefore produce non-Gaussian entangled states. It has been shown\cite{Fiurasek07} that in this case the entanglement distillation can be carried out using only interference
on beam splitters, balanced homodyne detection and conditioning on
the measurement outcomes. This was recently experimentally confirmed by successful demonstrations of distillation and purification protocols for single squeezed modes that suffered from de-Gaussifying noise \cite{Glockl06,Heersink06,Franzen06}.

This work experimentally demonstrates the first distillation
protocol for CV entangled states. Our protocol uses two copies of
phase-diffused\cite{Franzen06,Hage07}, therefore mixed, entangled
states exhibiting a positive, but non-Gaussian Wigner function. Our
protocol enhances the entanglement and purity of the decohered
states and represents a single step of an iterative Gaussification
scheme\cite{Browne03,Eisert04} that asymptotically converts any
input state into a Gaussian one. Moreover, if combined with a single
de-Gaussifying operation such as the recently demonstrated single-photon
subtraction from squeezed
beams\cite{Ourjoumtsev06,Neergaard06,Wakui07}, it would provide a
generic continuous variable  entanglement purification and
distillation scheme\cite{Browne03,Eisert04}, that is capable, for instance, of
suppressing the detrimental effect of losses in quantum state
transmission.
%


A schematic sketch of our protocol is shown in Figure \ref{setup}.
Two optical parametric amplifiers (OPAs) provided two continuous
wave light fields that carried squeezed states of light. Both
squeezed states were mixed with vacuum states on beam splitters with
50\% power reflectivity in order to prepare two copies of so-called
v-class \cite{DHFFS06} entangled states. Entanglement prepared in
this way is not the strongest possible but is effectual for this
proof of principle experiment. All four resulting  beams of the two
copies of entangled states were transmitted to two parties Alice (A)
and Bob (B) through four channels exhibiting independent phase
noise. The noisy channels were realized by quasi-random
electro-mechanical actuation of mirror positions in the beam paths
\cite{Franzen06, Hage07} in order to mimic the phase noise
introduced for example in optical fibres. The phase fluctuations
applied exhibited a Gaussian distribution, hence the standard
deviation of the noise $\spn$ provided a complete characterization
of its strength.

Alice and Bob each received two beams which they overlapped on a
balanced beam splitter (BS$_{\mathrm{D,A}}$ and BS$_{\mathrm{D,B}}$
in Fig.\,\ref{setup}). The mean phase was controlled such that the
initial quadrature phases lined up. The two output ports of each
beam splitter were connected to a total of four balanced homodyne
detectors (BHDs). Each of the BHDs could be set to observe an arbitrary quadrature,
in particular the amplitude quadrature X or the conjugate phase quadrature
P. The detectors BHD$_{\mathrm{T,A}}$ and
BHD$_{\mathrm{T,B}}$ were used to generate trigger events
discriminating the success or failure of  preparation a disitlled
state, while BHD$_{\mathrm{V,A}}$ and BHD$_{\mathrm{V,B}}$ were used
to independently verify the performance of the protocol. The latter
two were therefore not part of the actual protocol and could be
replaced by an arbitrary experiment. The fringe visibilities on all
four BHDs as well as on (BS$_\mathrm{D,A/B}$) were between 97.1\%
and 98.2\%. In our experiment both trigger detectors
BHD$_{\mathrm{T,A/B}}$ were set to measure the initially squeezed
amplitude quadratures $X_{T,A}$ and $X_{T,B}$, respectively. We
chose the trigger condition determining the success of the
distillation protocol in the form\cite{Fiurasek07}
\begin{equation}
|X_{T,A}+X_{T,B}|<Q, \label{trigger}
\end{equation}
where $Q$ was a certain threshold whose value could be tuned in
order to vary the selectivity of the protocol.

\begin{figure}[!t!]
\includegraphics[width=\linewidth]{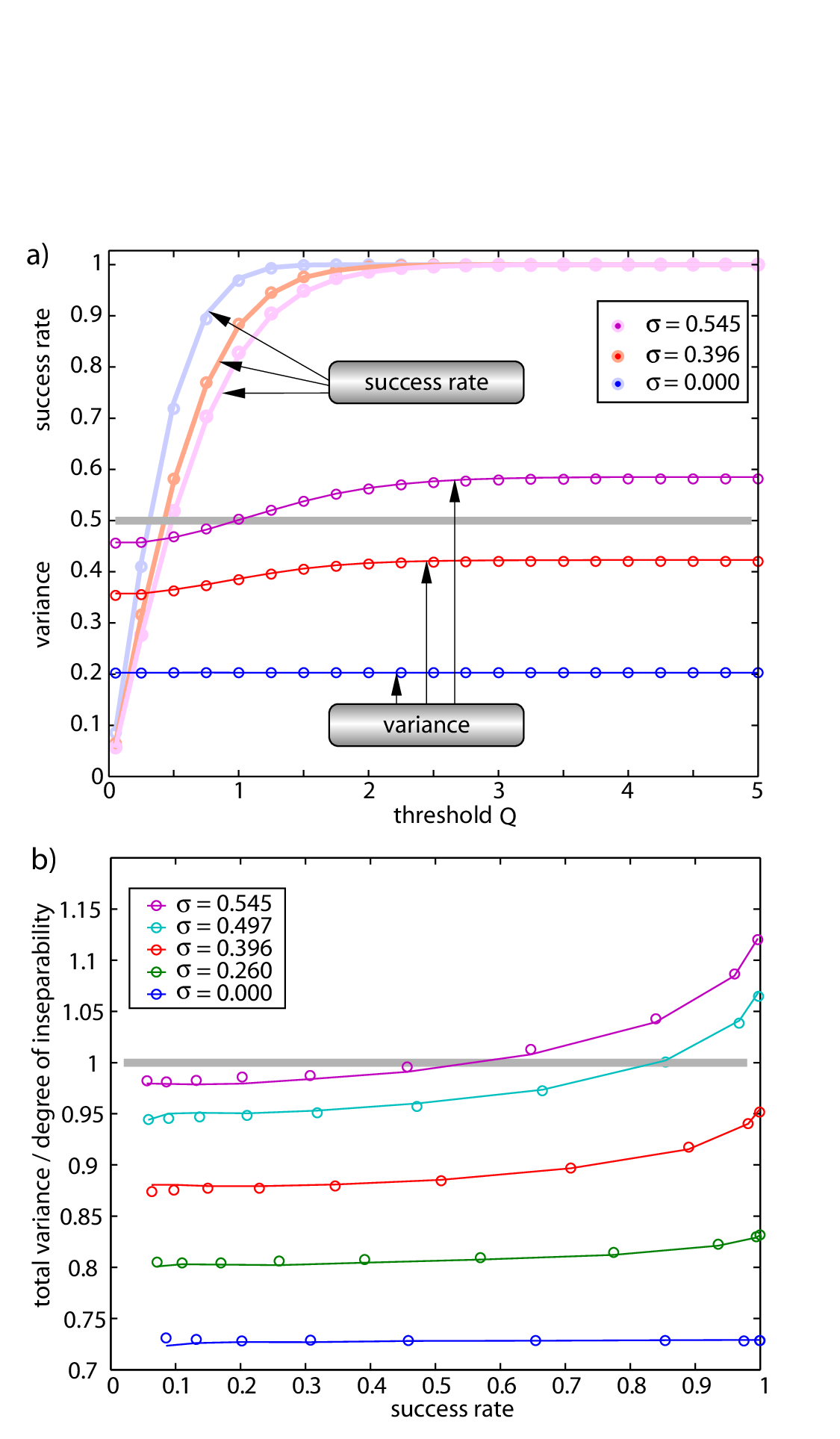}
\caption {\textbf{Nonlocal quadrature variance and total variance of the
distilled states.} (a) Variance of ($X_{V,A}+X_{V,B}$) and
corresponding success rates versus threshold value $Q$ for different
strengths of the phase fluctuations $\spn$. $\circ$~:~measurement,
--~:~simulation. (b) The total variance $\mathcal{I}$ plotted
against the success rate can be seen as measure for the performance
of our protocol. The~grey lines indicate the vacuum noise level or
separability boundary, respectively. For a success rate as high as
$0.5$ the protocol deploys nearly its full potential.
\label{varthresh}}
\end{figure}

Figure~\ref{varthresh}a) shows the variance of the nonlocal EPR-like
quadrature operator ($X_{V,A}+X_{V,B}$) of the successfully
distilled state and its corresponding probability of successful
preparation versus threshold value $Q$ for three different strengths
of the phase fluctuations $\spn$. The lower the threshold was set,
the more selectively our protocol worked. The variance of the orthogonal EPR-like
quadrature operator ($P_{V,A}-P_{V,B}$) was also reduced
(not shown). In this case the effect was rather small and limited by
the vacuum noise value of 1/2, because v-class\cite{DHFFS06}
entangled states were used in our experiment.

Figure~\ref{varthresh}b) shows the total variance $\mathcal{I}$ of
the distilled state versus its preparation success rate for five
different strengths of phase noise. The total variance quantifies
the quantum correlations between two modes and the degree of inseparability of the joint state (see Methods). If $\mathcal{I}<1$ then the state is not separable and therefore entangled\cite{DGCZ00}.  We can
see that the distillation and purification protocol reduces the
total variance $\mathcal{I}$ of the phase-diffused state. For the
two strongest levels of phase noise, the total variance shows no
nonclassical behaviour without the distillation protocol, but was
reduced below the unity boundary after applying the distillation protocol.
Lower values of $Q$ result in a stronger distillation effect but
also in a reduced success probability. It is promising to find that
for a success rate as high as 0.5 the protocol had nearly developed its
full potential. Note that the total variance of the entangled state
before applying phase noise was evaluated to be $\mathcal{I}=0.725$.

\begin{figure}[!t!]
\includegraphics[width=\linewidth]{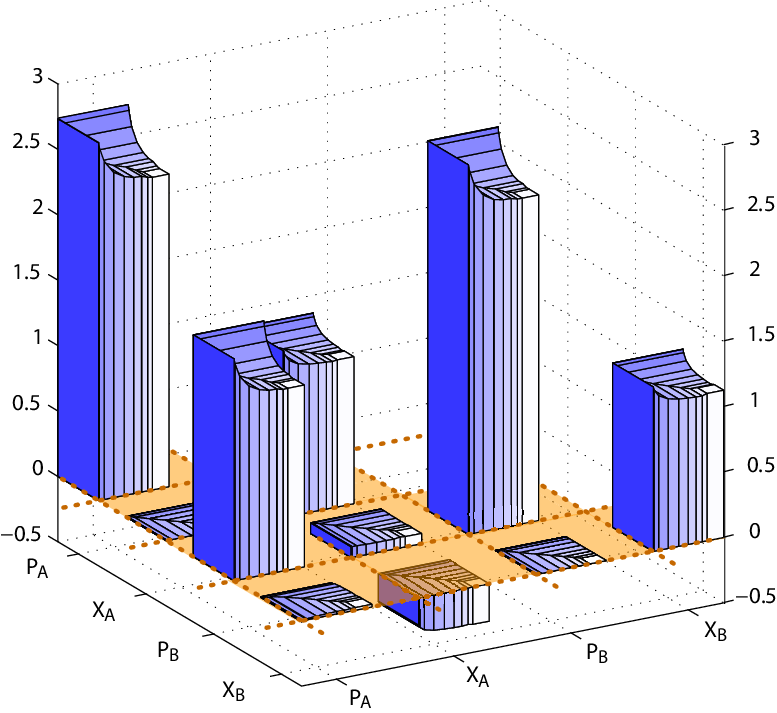}
\caption{\textbf{Reconstructed covariance matrices of distilled
entangled states.} Displayed is the lower left triangle of this
matrix containing its ten significant elements. The sub-bars
represent the results of our distillation protocol for different
threshold values $Q$ in the case of $\spn=0.497$. The base area of
each sub-bar represents the corresponding success rate. The vacuum reference for this plot is given by the unit matrix. \label{covmatrix}}
\end{figure}

Since the homodyne detectors can measure arbitrary quadrature
phases, we performed a tomographic reconstruction of the covariance
matrices of the distilled two-mode entangled states. We followed the
reconstruction procedure that is described in detail in our previous
work\cite{DHFFS06} and is based on joint measurements of quadrature
operators of modes $\Av$ and $\Bv$ for four different settings of
the BHDs. From this data we determined the most significant eight
of the ten independent parameters of the covariance matrix, namely the
variances of four quadratures $X_\Av$, $P_\Av$, $X_\Bv$, $P_\Bv$ and
covariances between all pairs of quadratures of Alice and Bob. The
intra-modal correlations were neglected, because the deviation from
zero was too small to be measured within the given accuracy of the
quadrature phase.
The resulting covariance matrices are plotted in Fig.~\ref{covmatrix} for ten different success rates. With
decreasing success rate the distillation becomes stronger as witnessed by the reduction of the quadrature
variances, i.e. diagonal elements of the covariance matrix. Moreover, the anti-correlation between $X_\Av$
and $X_\Bv$ was  slightly enhanced. Consequently, the squeezing of the nonlocal quadrature $X_\Av+X_\Bv$ was
enhanced by the distillation.

\begin{figure}[!t!]
\includegraphics[width=\linewidth]{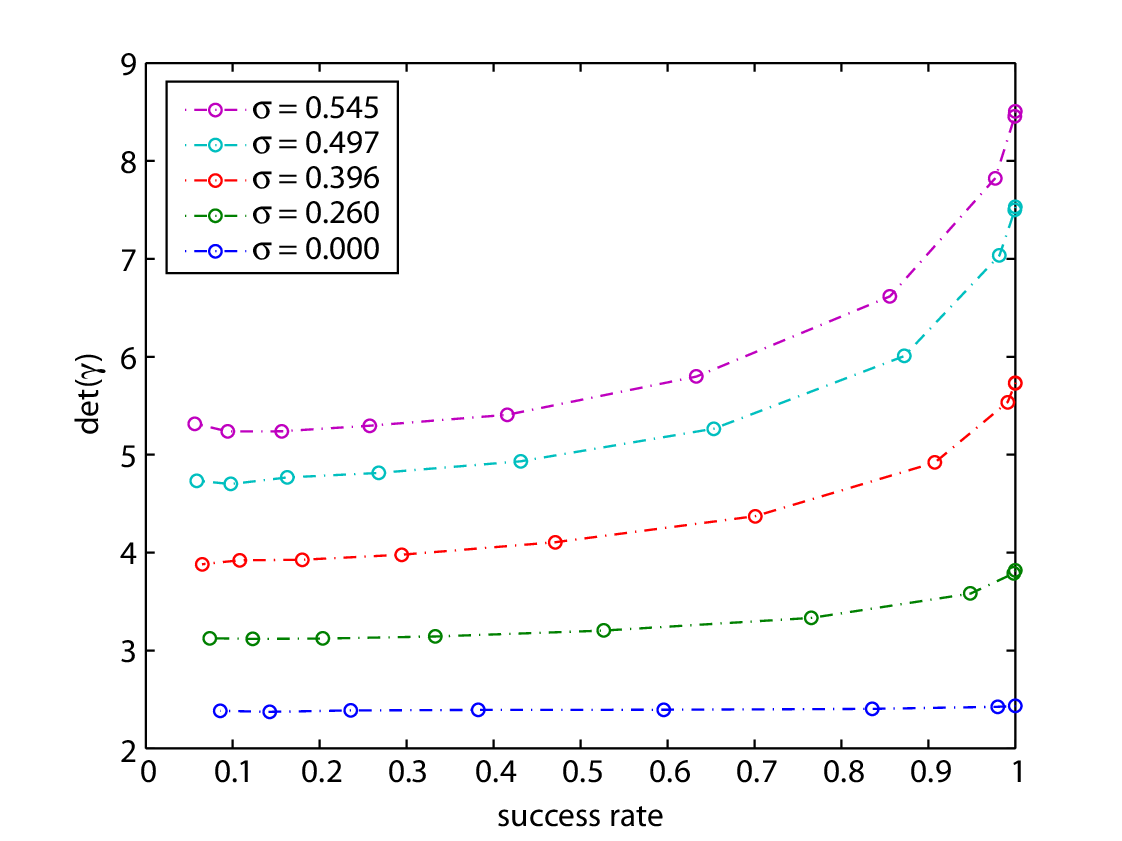}
\caption{\textbf{Determinant of covariance matrix of distilled
state} The determinant $D$ is plotted versus the success rate of the
distillation protocols for five different levels of phase noise. The
decrease of the determinant $D$ towards the left is an indication of
increased purity and Gaussification of the distilled state. }
\label{determinant}
\end{figure}

The determinant $D=\det(\gamma)$ is an important characteristic of a
covariance matrix $\gamma$. The purity $\mathcal{P}$ of a Gaussian
state $\rho$ defined as $\mathcal{P}= \mathrm{Tr}[\rho^2]$ is given
by $\mathcal{P}=1/\sqrt{D}$. In Fig.~\ref{determinant} we plot the
dependence of the determinant $D$ on the success rate of the
distillation protocol. The distillation reduces $D$ which is a very
strong indication of the increased purity of the (still slightly
non-Gaussian) distilled state, because a generalized Heisenberg
uncertainty relation implies that $D\geq 1$ and $D = 1$ holds only
for pure Gaussian states. The anticipation that the distilled states should not only show purification but also Gaussification was indeed confirmed by fitting Gaussian functions to the measurement data before and after the distillation protocol.
%

The protocol demonstrated here is able to distill entangled states
from a decohered ensemble that suffered from phase noise in the
transmission channel e.g. due to thermally excited refractive index
fluctuations or Brillouin scattering. Because of its quantum nature,
our protocol can counteract phase diffusion in cases where any
classical channel probing would fail, for example if the phase fluctuations are intensity dependent and arise from quantum fluctuations in the photon number of the transmitted
states.
Our protocol provides two \emph{open} ports that output the distilled
entangled states, and is therefore unconditionally useful for arbitrary
down-stream quantum information applications that involve the second
order moments of quadrature operators. An example is a teleportation
protocol that teleports Gaussian states\cite{Furusawa98}.
The reported entanglement
distillation, purification and Gaussification  protocol can be
iterated\cite{Browne03,Eisert04} and combined with already
experimentally demonstrated single-photon
subtraction\cite{Ourjoumtsev06,Neergaard06,Wakui07}, quantum
memory\cite{Julsgaard04} and entanglement
swapping\cite{Jia04,TYAF05} to build a continuous-variable quantum
repeater. Our experiment is thus an important enabling step towards
truly long-distance quantum communication with continuous variables.

\section*{METHODS}

\textbf{Generation and detection of squeezed states} Both optical
parametric amplifiers  were constructed from second order nonlinear
crystals ($\textrm{MgO:LiNbO}_3$) inside a degenerate doubly
resonant cavity \cite{Hage07} and were pumped with frequency doubled
laser beams from a monolithic solid state laser (Nd:YAG) operated at
$1064\,\textrm{nm}$. The OPAs emitted continuous wave squeezed beams
with about 4.5\,dB of squeezing and 8\,dB anti-squeezing at Fourier
frequencies around 7\,MHz.

All four output beams of the distillation protocol were measured by
balanced homodyne detectors (BHDs) each consisting of a balanced
beam splitter, a coherent local oscillator beam at 1064\,nm and two
photodiodes. The photocurrents produced by the photodiodes were
subtracted and amplified by home-made electronics. The signals from
the four BHDs were electronically mixed with a 7\,MHz local
oscillator. The intermediate signals were then anti-alias filtered
at 400\,kHz, synchronously sampled with 1\,MHz and post-processed to
accomplish and verify the distillation protocol.

\textbf{Entanglement detection and quantification} The entanglement
of the distilled two-mode squeezed output state can be conveniently
certified by the Duan criterion\cite{DGCZ00}. This criterion
involves the sum of variances of two nonlocal, EPR-like quadrature operators,
the so-called total variance, $\mathcal{I}=\langle(\Delta
X_{\mathrm{+,V}})^2\rangle+\langle (\Delta
P_{\mathrm{-,V}})^2\rangle$, where $X_{\mathrm{+,V}}=X_\Av+X_\Bv$
and $P_{\mathrm{-,V}}=P_\Av-P_\Bv$. The state is entangled if
$\mathcal{I}<1$, where the vacuum state corresponds to a value of
$\mathcal{I}=1$. The total variance $\mathcal{I}$ provides a
quantitative measure of the quantum correlations between the
quadratures of the distilled states. For symmetric two-mode squeezed
Gaussian states $\mathcal{I}$ fully quantifies the entanglement of
the state\cite{Giedke03}. It has been shown
numerically\cite{Fiurasek07} that even for the non-Gaussian phase
diffused squeezed states there is a one-to-one correspondence
between the reduction of the total variance and the increase of the
logarithmic negativity, which is a computable entanglement
measure\cite{Vidal02}.

\vspace*{2mm}

\textbf{Acknowledgements} J.F. acknowledges the financial support of
the Future and Emerging Technologies (FET) programme within the
Seventh Framework Programme for Research of the European Commission,
under the FET-Open grant agreement COMPAS, number 212008 and from
MSMT under projects Center of Modern Optics (LC06007) and
Measurement and Information in Optics (MSM6198959213). R.S.
acknowledges financial support from the Deutsche
Forschungsgemeinschaft (DFG), Project No. SCHN 757/2-1.

\vspace*{2mm}

\textbf{Competing Interests} The authors declare that they have no competing financial interests.

\vspace*{2mm}

\textbf{Correspondence} Correspondence and requests for materials should be addressed to
R. Schnabel (email: roman.schnabel@aei.mpg.de).

\end{document}